# Detection and classification of DDoS flooding attacks by machine learning method

Dmytro Tymoshchuk [1,*,†], Oleh Yasniy [1,†], Mykola Mytnyk[1,†], Nataliya Zagorodna[1,†] and Vitaliy Tymoshchuk[1,†]

[1] *Ternopil Ivan Puluj National Technical University, Ruska str. 56, Ternopil, 46001, Ukraine*

**Abstract**
This study focuses on a method for detecting and classifying distributed denial of service (DDoS) attacks, such as SYN Flooding, ACK Flooding, HTTP Flooding, and UDP Flooding, using neural networks. Machine learning, particularly neural networks, is highly effective in detecting malicious traffic. A dataset containing normal traffic and various DDoS attacks was used to train a neural network model with a 24-106-5 architecture. The model achieved high Accuracy (99.35%), Precision (99.32%), Recall (99.54%), and F-score (0.99) in the classification task. All major attack types were correctly identified.
The model was also further tested in the lab using virtual infrastructures to generate normal and DDoS traffic. The results showed that the model can accurately classify attacks under near-real-world conditions, demonstrating 95.05% accuracy and balanced F-score scores for all attack types. This confirms that neural networks are an effective tool for detecting DDoS attacks in modern information security systems.

## 1. Introduction

Distributed denial of service (DDoS) attacks are one of the most serious threats to network security. These attacks cause significant system disruptions by flooding the system with malicious traffic [1]. Among the various DDoS techniques, Flooding attacks, such as SYN Flooding, ACK Flooding, HTTP Flooding, and UDP Flooding, are particularly difficult to neutralise due to their ability to mimic legitimate traffic. These attacks drain server resources, making it unavailable to legitimate users.

Machine learning (ML) is one of the key technologies increasingly being implemented in various fields of science and technology due to its ability to automate processes, analyze large amounts of data, and make highly accurate predictions. In medicine, ML is used to diagnose diseases, analyze medical images, develop personalized treatment plans, and predict the spread of infectious diseases [2]. In the financial sector, machine learning allows for assessing credit risk, detecting fraud, optimizing investment portfolios, and automating trading algorithms [3]. In the automotive industry, ML underpins the development of autonomous vehicles that analyze sensor data to make real-time decisions and predict vehicle maintenance [4]. In materials science, machine learning allows predicting material properties [5,6,7]. In particular, ML minimizes the need for expensive and time-consuming experiments.

In cybersecurity, machine learning has become an important tool for detecting and preventing various threats. Traditional methods, such as rule-based and statistical approaches, often cannot detect sophisticated attacks. ML allows for more efficient analysis of network traffic, detection of anomalies, and classification of malicious traffic, making these methods indispensable for modern information security systems [8,9]. Neural networks are a subset of machine learning techniques known for their ability to detect complex nonlinear relationships.

The aim of this study is to develop and evaluate an effective neural network for DDoS detection and classification. A dataset containing normal traffic and traffic from different Flooding attacks (SYN, ACK, HTTP, and UDP Flooding) was used to train and evaluate the neural network model. The main stages of this work include the development of a robust neural network model for DDoS detection, analyzing its performance for different types of attacks, and practical testing of the neural network performance on traffic generated in a laboratory environment under conditions close to real-world conditions.

## 2. Methods

### 2.1. Dataset description

The dataset used in this study is specifically designed to detect DDoS flooding attacks and is obtained from [10,11]. It includes two categories of traffic: normal traffic, which represents legitimate user activity, and malicious traffic generated by different Flooding attacks.

Normal traffic is network traffic that does not contain malicious activity and corresponds to the standard behavior of users and devices on the network. Such traffic includes legitimate requests, data transfers between users and servers, and other typical network operations that occur during the normal operation of network systems. In the context of DDoS detection, normal traffic is a benchmark for comparison with abnormal traffic that may indicate an attack.

SYN Flooding is a type of DDoS attack aimed at exhausting the resources of a server or network device's resources by sending many requests to establish a TCP connection [12]. Under normal conditions, a TCP connection is established through a three-step process where the client sends a SYN packet to the server; after that, the server responds with a SYN-ACK, and the client completes the process by sending an ACK packet. However, in a SYN Flooding attack, the attacker sends many SYN packets (in many cases from a spoofed IP address) but does not respond to the SYN-ACK packets the server receives in return. This causes the server to keep half-open connections, wasting its resources on maintaining them. As a result, the server becomes overloaded and unable to process new legitimate connection requests, resulting in a denial of service for legitimate users. SYN Flooding is one of the most common and difficult attacks to detect because its packets look like legitimate requests.

ACK Flooding is a DDoS attack that uses many ACK packets to overload the target system [13]. ACK packets are part of the normal data transfer process in the TCP protocol and acknowledge receipt of a data packet from the sender. In a typical scenario, after data is transmitted between two devices, the receiver sends an ACK packet to the sender to confirm that the data was successfully received. In the case of ACK Flooding, an attacker sends many ACK packets to the target server or network device. These packets do not correspond to the connection or previously transmitted data. The attack aims to overwhelm the server by processing many invalid ACK packets, thereby depleting its resources, such as CPU time and memory. Due to the constant flow of ACK packets, the server is forced to spend significant resources on processing them, which can lead to a decrease in performance or a complete cessation of service to legitimate users. Like other DDoS attacks, ACK Flooding is difficult to detect because individual ACK packets are not malicious and look like normal traffic. However, their massive number and the lack of a suitable connection make the attack effective and lead to server overload.

HTTP Flooding is a DDoS attack that aims to exhaust web server resources by sending many HTTP requests [14]. In this case, attackers use the HTTP protocol to communicate between web browsers and servers to overload the target website or web application. In HTTP Flooding, attackers send requests that mimic legitimate web traffic to the target server. These requests can be for various website resources, such as HTML pages, images, or other media files. The attack aims to consume available server resources, such as network bandwidth, CPU time, and RAM. HTTP Flooding can significantly impact a website or web application, especially if the attack is large-scale. Due to the heavy load, the server can slow down or even stop functioning completely, making the website inaccessible to legitimate users. Since HTTP Flooding uses normal web traffic, it is difficult to distinguish it from legitimate requests, making it difficult to recognize and block the attack.

UDP Flooding is a DDoS attack that uses many UDP packets to overload a target server or network device [15]. UDP (User Datagram Protocol) is a data transmission protocol that does not check packet delivery and does not establish a connection before sending data. In the case of UDP Flooding, attackers send many UDP packets to random ports on the target server or network device. When the server receives these packets, it tries to process them, including checking the incoming data and attempting to respond to requests if necessary. As a result, the server spends resources processing and responding to large volumes of spoofed requests. This leads to an overload of its network bandwidth and CPU resources, which can significantly slow down or stop the server's normal operation. UDP Flooding can also affect the network infrastructure by flooding communication channels with large data. Like other DDoS attacks, UDP Flooding can be difficult to detect and block because UDP packets are not malicious, and the attack uses a legitimate network protocol.

The dataset has already been pre-processed to make it suitable for neural network training. The dataset is divided into three parts to train the neural network: training, test, and validation samples. The total sample size was 38413. Of these, 16619 records were normal traffic, 3556 were SYN Flooding, 7562 were ACK Flooding, 1044 were HTTP Flooding, and 9632 were UDP Flooding.

To ensure effective training and evaluation of the model, 70% of the data were randomly selected for the training set, the largest share of the data. This part was used to train the model, i.e. to adjust its parameters based on the available data. The validation sample comprised 15% of the total data. It was used to check the quality of the model and the settings of its hyperparameters. This allowed us to avoid overfitting, i.e. a situation where the model works well on training data but performs poorly on new, unknown data. The remaining 15% of the data was reserved for the test sample, which was used after the model was trained. Testing allowed us to evaluate the final performance of the model on new data that was not involved in the training or validation process. It allowed us to determine its generalization capability.

## 2.2. Neural network model

A neuron in neural networks is the basic element that mimics the behaviour of a biological neuron. Its main function is to receive signals at the input, process them, and transmit the results to the output. The mathematical model of the neuron is described by the following equation [16]:

$$y = \varphi\left(\sum_{i=1}^{n} \omega_i \cdot x_i + b\right), \tag{1}$$

where $x_i$ are the input values, $\omega_i$ are the weights associated with each input, $b$ is the bias that allows the neuron to better adapt to the data, $b$ is the activation function, $n$ is the number of input signals or the number of input features, $y$ is the output of the neuron. Each neuron receives input signals represented as a set of values $x_1, x_2, \ldots, x_n$. These values are weighted according to their weights $\omega_1, \omega_2, \ldots, \omega_n$, which are adjusted during model training. The sum of the weighted inputs is then passed to the activation function $\varphi$.

In this paper, a neural network with the 24-106-5 architecture is built. Figure 1 shows the architecture of a multilayer perceptron with one hidden layer and one output layer.

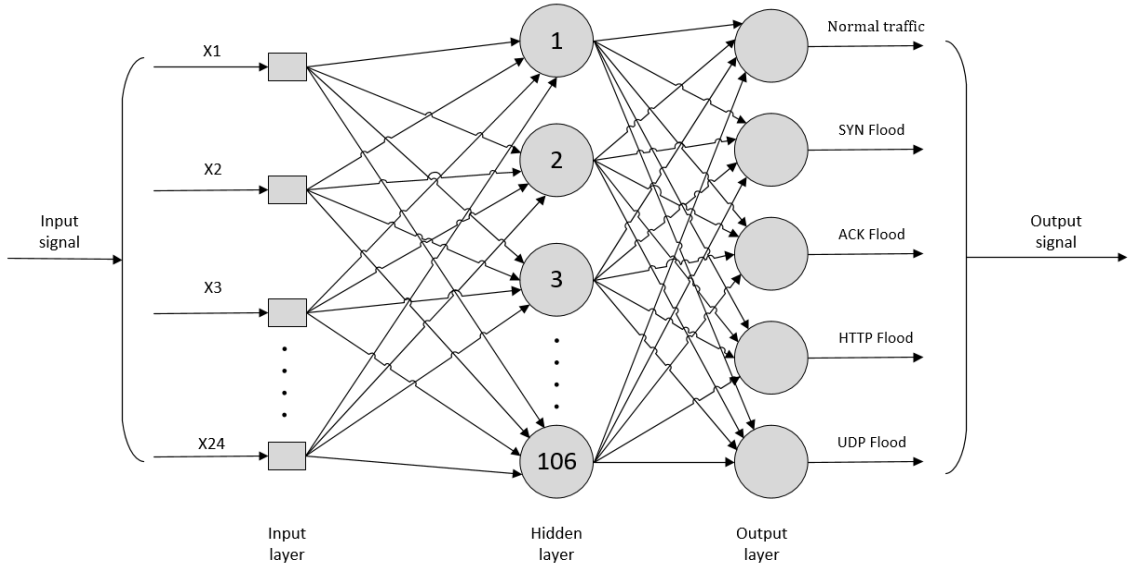

**Figure 1:** Architecture of the 24-106-5 feed-forward neural network

A neural network consists of three layers. The input layer contains 24 nodes, corresponding to the number of features in the input data; each neuron receives values from the data set and passes them on to the next layer. The hidden layer comprises 106 neurons activated using the tanh (hyperbolic tangent) activation function. The function is described as:

$$tanh(x) = \frac{e^x - e^{-x}}{e^x + e^{-x}}, \qquad (2)$$

where $x$ is an input value or a weighted sum of input signals for a particular neuron, $e$ is a mathematical constant known as the Euler number.

The tanh function maps input values into a range from -1 to 1, which allows the model to learn more efficiently as it reduces the problem of vanishing gradients compared to other activation functions. The output layer contains 5 neurons, corresponding to the number of classes in the classification task, including normal traffic and four types of DDoS Flooding attacks. To activate the neurons in this layer, we used the softmax activation function. The function is described as:

$$\sigma(z_i) = \frac{e^{z_i}}{\sum_{j=1}^{n} e^{z_j}}, \qquad (3)$$

where $z_i$ is a real number that reflects the 'strength' of the signal for class $i$ before applying softmax, $i$ is a fixed index for calculating the probability of a particular class, $n$ defines the number of possible categories for classification.

This function maps the neurons' output values into probabilities belonging to each class, where the sum of probabilities for all classes is 1. This allows the model to provide probabilities for each possible outcome, which is convenient for the classification task.

This neural network architecture allows for the effective detection and classification of various DDoS attacks using input data consisting of various network traffic characteristics.

## 2.3. Model evaluation

A confusion matrix was built to evaluate the neural network's effectiveness in detecting and classifying DDoS attacks (Table 1). The model evaluation is based on four main categories of classification results: True Positive (TP), True Negative (TN), False Positive (FP), and False Negative (FN).

**Table 1**
Confusion matrix

| True label | Predicted label | |
|---|---|---|
| | Normal traffic | DDoS traffic |
| Normal traffic | TN | FP |
| DDoS traffic | FN | TP |

True Positive (TP) represents the number of times the model correctly identified an attack. True Negative (TN) represents the number of times the model correctly identified normal traffic. False Positive (FP) represents the number of times the model incorrectly identified normal traffic as an attack. False Negative (FN) shows the number of cases when the model failed to recognize an attack and classified the data as normal traffic.

Several key performance indicators are calculated based on the TP, TN, FP, and FN values: Accuracy, Precision, Recall, Specificity, and F-score.

In terms of classifying normal traffic and DDoS traffic, accuracy shows the overall efficiency of the model in classifying these two types of traffic:

$$Accuracy = 100\% \cdot \frac{\sum(TP + TN)}{\sum(TP + TN + FP + FN)}, \qquad (4)$$

If the model correctly identifies the majority of samples as either normal traffic or a DDoS attack, it will have high accuracy.

Precision in the context of DDoS detection is the proportion of correctly classified traffic as a DDoS attack among all samples that are classified as attacks:

$$Precision = 100\% \cdot \frac{\sum(TP)}{\sum(TP + FP)}, \qquad (5)$$

A high score means that the model rarely mistakes normal traffic for a DDoS attack, meaning that the number of False Positive results is low. This is important in real-world networks, where false alarms can lead to unnecessary blocking of legitimate traffic.

Recall (Sensitivity) shows how well the model detects real DDoS attacks. It is the proportion of correct attack classifications among all the real attacks present in the dataset:

$$Recall = 100\% \cdot \frac{\sum(TP)}{\sum(TP + FN)}, \qquad (6)$$

A high score means that the model effectively detects most or all real DDoS attacks while minimizing the number of missed attacks (False Negative). This is critical because missed attacks can go undetected, allowing attackers to cause damage to the system.

Specificity measures how well the model identifies normal traffic and distinguishes it from DDoS attacks:

$$Specificity = 100\% \cdot \frac{\sum(TN)}{\sum(TN + FP)}, \qquad (7)$$

A high score means that the model correctly classifies most normal traffic samples as not DDoS, reducing the number of False Positive results. It indicates how well the model protects legitimate traffic from false blocking.

The F-score in the case of classifying normal traffic and DDoS attacks provides a balanced assessment between Precision and Recall. It allows you to evaluate the overall performance of the model, taking into account both the model's ability to minimize False Positive results (high Precision) and detect genuine attacks (high Recall):

$$F - score = \frac{2 \cdot Recall \cdot Precision}{Recall + Precision}, \qquad (8)$$

A high F-score indicates that the model performs well in detecting real attacks and avoiding false alarms, which is key to reliable network protection.

These metrics provide a comprehensive assessment of the neural network's performance to classify network traffic and determine how effectively it detects and recognizes different types of DDoS attacks in combination with normal traffic.

## 3. Results and discussion

### 3.1. Detection performance

In this work, we used a neural network with a 24-106-5 architecture to detect and classify DDoS attacks such as SYN Flooding, ACK Flooding, HTTP Flooding, and UDP Flooding along with normal traffic. The confusion matrix of the model is shown in Table 2.

**Table 2**
Confusion matrix of the neural network 24-106-5

| True label | Predicted label | | | | |
|---|---|---|---|---|---|
| | Normal traffic | SYN Flooding | ACK Flooding | HTTP Flooding | UDP Flooding |
| Normal traffic | 2471 | 15 | 3 | 2 | 2 |
| SYN Flooding | 10 | 523 | - | - | - |
| ACK Flooding | 1 | - | 1133 | - | - |
| HTTP Flooding | 1 | - | - | 156 | - |
| UDP Flooding | 3 | - | - | - | 1442 |

Most of the traffic samples are classified correctly, but it is noticeable that the largest classification errors occur between normal traffic and SYN Flooding. In some cases, the difference between normal connection establishment and SYN Flooding can be small, making it difficult for the model to distinguish between these types of traffic accurately. This also explains why the model sometimes confuses the two types of traffic, as SYN packets are used both in the normal connection establishment process and during an attack. Therefore, the similarity like requests between normal traffic and SYN Flooding may be the main reason for the increase in classification errors between them.

Table 3 shows the performance of the neural network in the task of detecting and classifying DDoS attacks.

**Table 3**
Performance indicators of the neural network 24-106-5

| Performance indicator | Normal traffic | | | | |
|---|---|---|---|---|---|
| | All DDoS Flooding | SYN Flooding | ACK Flooding | HTTP Flooding | UDP Flooding |
| Accuracy (%) | 99.35 | 99.17 | 99.88 | 99.88 | 99.87 |
| Precision (%) | 99.32 | 97.21 | 99.73 | 98.73 | 99.86 |
| Recall (%) | 99.54 | 98.12 | 99.91 | 99.36 | 99.79 |
| Specificity (%) | 99.11 | 99.39 | 99.87 | 99.91 | 99.91 |
| F-score | 0.99 | 0.97 | 0.99 | 0.99 | 0.99 |

The overall accuracy rate is 99.35%, which indicates that the model is highly effective in correctly classifying different types of traffic. The model also shows a high Precision of 99.32%, which indicates a minimal number of False Positive results. Recall, which reflects the model's ability to detect real attacks, reaches 99.54%, which means that the model almost never misses real attacks. The Specificity, which indicates the model's ability to identify normal traffic correctly, is 99.11%, which confirms the model's high ability to avoid misclassifying normal traffic as an attack. The F-score is 0.99, emphasizing the model's balance in detecting attacks and minimizing false alarms. Among the individual attack types, ASK Flooding and UDP Flooding are the easiest for the model to detect, with performance scores above 99% for all metrics. At the same time, the performance for SYN Flooding is somewhat lower, especially in Precision and F-score, which may be due to the peculiarities of this type of attack. Overall, the model shows a high level of performance, making it a reliable tool for detecting various types of DDoS attacks.

### 3.2. Practical testing of a neural network

The machine learning approach to detecting DDoS traffic and normal traffic involves several key steps (Figure 2).

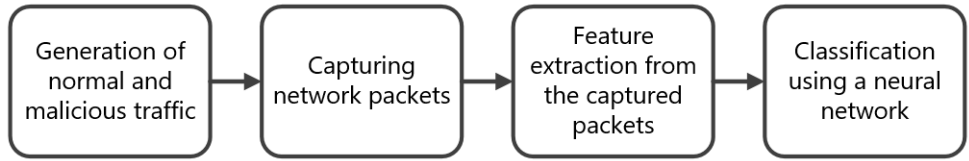

**Figure 2:** Steps of the malicious traffic detection approach

To implement this approach, a special network infrastructure was created (Figure 3).

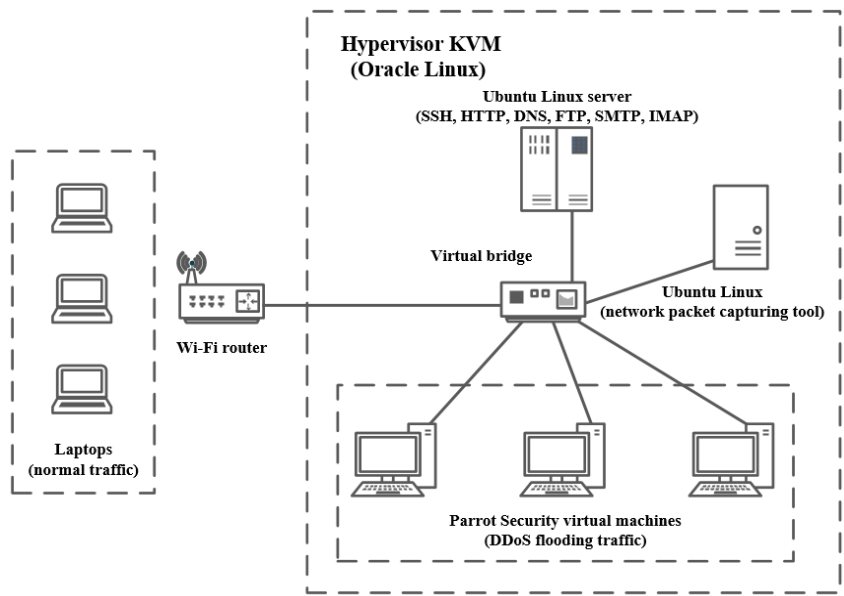

**Figure 3:** Lab network infrastructure

The basis of this infrastructure is a KVM hypervisor installed on Oracle Linux, which provides virtualization and supports the operation of virtual machines. Communication between the virtual machines and the physical network occurs through a virtual bridge connected to a Wi-Fi router. The Ubuntu Linux server deploys network services such as SSH, HTTP, DNS, FTP, SMTP, and IMAP, which are used to create a realistic network environment. Virtual machines with Parrot Security were launched to generate DDoS traffic to test the DDoS detection algorithms. At the same time, a network packet capture tool is running on Ubuntu Linux, collecting all network traffic for further analysis. Normal traffic, used to simulate normal network activity, is generated by laptops connected to the same network via Wi-Fi. This configuration allows you to collect the necessary data to test the effectiveness of the neural network in realistic conditions by simulating different types of traffic.

We used tcpdump to capture network packets and save them to a file in the pcap format. Parrot Security created the DDoS traffic using Metasploit, an effective tool for carrying out network attacks, including SYN Flooding, ACK Flooding, HTTP Flooding, and UDP Flooding.

Special software was developed in the Python programming language to extract features from captured network packets. This software allows for the automatic processing and analysis of large network data, highlighting key features that allow for further traffic classification.

The total network traffic records created in the lab environment was 10564. Of these, 3845 records were normal traffic, 1721 were SYN Flooding, 2203 were ACK Flooding, 980 were HTTP Flooding, and 1815 were UDP Flooding. The confusion matrix of the model is shown in Table 4.

**Table 4**
Confusion matrix of neural network 24-106-5 in practical testing

| True label | Predicted label | | | | |
| --- | --- | --- | --- | --- | --- |
| | Normal traffic | SYN Flooding | ACK Flooding | HTTP Flooding | UDP Flooding |
| Normal traffic | 3467 | 126 | 101 | 64 | 87 |
| SYN Flooding | 56 | 1665 | - | - | - |
| ACK Flooding | 34 | - | 2169 | - | - |
| HTTP Flooding | 25 | - | - | 955 | - |
| UDP Flooding | 29 | - | - | - | 1786 |

This confusion matrix shows the results of testing a neural network to detect and classify DDoS attacks from traffic generated in a lab environment. The neural network detected 3467 cases of normal traffic correctly. Still, several errors were made when SYN Flooding, ACK Flooding, HTTP Flooding, and UDP Flooding were mistakenly identified as normal traffic. Also, for SYN Flooding, the network correctly predicted 1665 cases but made 56 mistakes, identifying it as normal traffic. In the case of ACK Flooding, the network correctly predicted 2169 samples but made 34 errors. The network accurately classified 955 HTTP Flooding cases, making minor errors with this category, and correctly predicted 1786 UDP Flooding cases. The network generally does a good job of classifying DDoS attacks, but there are several errors, especially when SYN and ACK Flooding are classified as normal traffic.

Table 5 shows the neural network's performance in detecting and classifying DDoS attacks on traffic generated in the laboratory environment.

**Table 3**
Performance indicators of the neural network 24-106-5 in practical testing

| Performance indicator | Normal traffic | | | | |
|---|---|---|---|---|---|
| | All DDoS Flooding | SYN Flooding | ACK Flooding | HTTP Flooding | UDP Flooding |
| Accuracy (%) | 95.05 | 96.57 | 97.66 | 98.02 | 97.83 |
| Precision (%) | 94.56 | 92.96 | 95.55 | 93.71 | 95.35 |
| Recall (%) | 97.85 | 96.74 | 98.45 | 97.44 | 98.40 |
| Specificity (%) | 90.16 | 96.49 | 97.16 | 98.18 | 97.55 |
| F-score | 0.96 | 0.95 | 0.97 | 0.95 | 0.97 |

The overall Accuracy for all DDoS attacks is 95.05%, which indicates that the model can classify both normal traffic and different types of attacks correctly. Precision, which determines the percentage of correct positive predictions among all predicted positive cases for all attacks, is 94.56%. Recall, which indicates how well the model detects all positive cases, has the highest performance among the other metrics. The overall Recall is 97.85%. Specificity, which shows how well the model avoids false positives, is also quite high. The overall score for all attacks is 90.16%. For all attacks, the F-score, the harmonic mean between Precision and Recall, is 0.96.

Thus, the neural network demonstrates high efficiency in detecting and classifying DDoS attacks on traffic generated in the laboratory environment, showing good results for all major metrics for each type of attack.

## 4. Conclusions

As a result of the study, the neural network showed high efficiency in detecting and classifying DDoS attacks. The overall accuracy is 95.05%, and the Precision, Recall, and Specificity values are high for all types of attacks, indicating the model's reliability. For all DDoS attacks, the overall F-score is 0.96, indicating that the model is highly balanced. This means that the model effectively detects genuine attacks without generating many false alarms.